\documentstyle[prb,aps,graphicx,multicol]{revtex}

\begin{document}

\title{Spin-accumulation and Andreev-reflection in a mesoscopic
ferromagnetic
  wire}

\author{W. Belzig}
\address{Department of Applied Physics and Delft Institute of
Microelectronics and Submicrontechnology,\\ Delft University of
Technology, Lorentzweg 1, 2628 CJ Delft, The Netherlands}

\author{Arne Brataas}
\address{Lyman Laboratory of Physics, Harvard University, Cambridge, MA
02138}

\author{Yu. V. Nazarov and Gerrit E. W. Bauer}
\address{Department of Applied Physics and Delft Institute of
Microelectronics and Submicrontechnology,\\ Delft University of
Technology, Lorentzweg 1, 2628 CJ Delft, The Netherlands}

\maketitle

\begin{abstract}
  The electron transport though ferromagnetic metal | superconducting hybrid
  devices is considered in the non-equilibrium Green's function formalism in
  the quasiclassical approximation. Attention if focused on the limit in which
  the exchange splitting in the ferromagnet is much larger than the
  superconducting energy gap. Transport properties are then governed by an
  interplay between spin-accumulation close to the interface and Andreev
  reflection at the interface. We find that the resistance can either be
  enhanced or lowered in comparison to the normal case and can have a
  non-monotonic temperature and voltage dependence. In the non-linear voltage
  regime electron heating effects may govern the transport properties, leading
  to qualitative different behaviour than in the absence of heating effects.
  Recent experimental results on the effect of the superconductor on the
  conductance of the ferromagnet can be understood by our results for the
  energy-dependent interface resistance together with effects of
  spin-accumulation without invoking long range pairing correlations in the
  ferromagnet.
\end{abstract}


\begin{multicols}{2}
\section{Introduction}
\label{sec:intro}

Much theoretical and experimental work has addressed the effect of a
superconductor (S) in proximity to a normal metal (N) on the transport
properties during the last years, see Ref.~\onlinecite{Belzig99:1251}
and references therein for a review. Most experimental results can be
explained in the framework of the quasiclassical theory of
superconductivity accounting for a ``long range'' proximity effect
with a coherence length $\xi=(\hbar D/2k_{\text{B}}T)^{1/2}$, where
$D$ is the diffusion coefficient of the normal metal and $T$ is the
temperature.  On the other hand, applications of the quasi-classical
theory to transport in heterostructures containing ferromagnets (F)
are still scarce. In contrast to normal metals the presence of a
strong exchange field in the ferromagnet leads to a strong difference
in the energy dispersions for the two spin bands.  However, long-range
coherence in normal metals requires spin degenerate bands close to the
Fermi energy, since singlet superconductivity couples quasiparticles
of different spins by Andreev reflection. The consequence of the
exchange field energy $h_{\text{xc}}$ is a strong decoherence of
quasiparticles belonging to the different spin bands.  Typically the
superconducting energy scale $\Delta$ is smaller than $h_{\text{xc}}$
by several orders of magnitude for (Al, Nb) {\em vs.} (Fe, Ni, and
Co), respectively.  Thus, the proximity effect in ferromagnetic metals
is negligible and a ferromagnet in contact to a superconductor may be
considered as an {\em incoherent} metal coupled to the superconductor.
In this case all changes induced by the contact to a superconductor
depend on the properties of the interface itself. This is accomplished
by the effect of spin accumulation\cite{Levy94:367,Brataas99:93},
which requires no phase coherence in the ferromagnet and can therefore
have a much longer range than the proximity effect.  The main purpose
of this paper is to study the mutual influence of resistance changes
by spin accumulation and interface properties.

Recently heterostructures of ferromagnets and superconductors have been
experimentally realized and
investigated.\cite{Upadhyay98:3247,Petrashov99:3281,Giroud98:R11872,Lawrence96}
Several unusual phenomena have been unveiled.  The experimental results in
point contact geometries\cite{Upadhyay98:3247} can be explained by the
reduced, bias-dependent transparency of the interface due to spin-dependent
band mismatch between the normal metal and the
ferromagnet.\cite{Upadhyay98:3247,deJong95:1657} The experimental results in
diffusive nanostructured
samples\cite{Petrashov99:3281,Giroud98:R11872,Lawrence96} are more intriguing.
The measured conductance changes on the ferromagnetic side can be positive and
negative at the superconducting transition with amplitudes much larger than
anticipated.  The sign and the amplitude of the changes appear to depend
strongly on the ferromagnetic - superconductor interface transparency. It has
been conjectured that a strong mutual influence of the superconductors and
ferromagnetic conductors and a penetration of the superconducting order
parameter into the ferromagnet over distances many times longer than expected
from the above estimates might explain the
observations.\cite{Petrashov99:3281,Giroud98:R11872,Lawrence96}

Some effects of the interplay between spin accumulation and Andreev reflection
in diffuse systems have been discussed in Refs.~\onlinecite{Falko99:532}.
Since the spin-current into a superconductor vanishes at sufficiently low bias
and temperature, a non-equilibrium spin-accumulation builds up on the
ferromagnetic side in order to conserve the spin-currents. The
spin-accumulation causes an additional boundary resistance which is of the
order of the resistance of the ferromagnetic wire of a length of the spin-flip
diffusion length. Therefore the resistance of the F-S system should be always
larger than that of the F-N system, in contradiction with many of the
experimental observations. A possible reason for this apparent failure is the
neglect of changes in the interface resistance in the transition from F-S to
F-N.  Previous theories took into account only perfectly ballistic interfaces
for which the resistance is determined purely by the matching of the adjacent
Fermi surfaces. The interface resistance and its modulation are of the order
of the Sharvin resistance, which is negligible compared to the total one.
However, in the sputtered samples with relatively large contact
areas\cite{Petrashov99:3281,Giroud98:R11872,Lawrence96} the interface can
contribute significantly, especially when differences of resistances below and
above the superconducting transition temperatures are considered.  Other
transport phenomena in ferromagnet-superconductors systems have been studied
in Ref. \onlinecite{Kadigrobov99:14593}.

It has been speculated that the triplet component of the order parameter
induced by the fluctuations of the spin-orbit scattering potentials is
essential in mesoscopic junctions.\cite{Zhou9906177} Neglecting magnetic
impurities and spin-orbit coupling the superconducting order parameter is a
spin singlet. However, magnetic impurities or spin-orbit coupling, induce a
fluctuating spin triplet component with zero average.  The triplet component
is `long-range' coherent in the ferromagnet since it couples electrons and
holes with the same spin and the exchange field in the ferromagnet does not
play a role. However, the contribution to the conductivity from the triplet
fluctuations is only relevant when the fluctuations are relatively large which
is only the case when the conductance is close to the quantum conductance.
The experimental samples\cite{Petrashov99:3281,Giroud98:R11872,Lawrence96}
have a much larger conductance, and we do not expect that such mesoscopic
fluctuations play an important role.

None of the above-mentioned theories can explain the recent experimental
results. This makes it necessary to study the properties of the contact
between the ferromagnet and the normal metal in more detail and to account for
a possible spin accumulation and heating effects in the ferromagnet including
all possible interfaces between the ferromagnets and the superconductors.  In
particular we will go beyond the assumptions of a perfect transparent metallic
interface\cite{Falko99:532} and discuss its influence on the observed
conductance changes below the superconducting transition temperature and for
bias voltages less than the superconducting gap. We will in this work
radically disregard the proximity effect. Therefore our results only apply to
ferromagnets with $h_{\text{xc}}\gg\Delta$, which is {\em e.g.} the case for
the magnetic transition metals (Fe, Co, and Ni) in conjunction with
superconducting metals like Nb and Al. This assumption is supported by the
experimental fact that FS interferometers show no phase-periodic oscillations
down to the level of $0.1 e^2/h$ in strong
ferromagnets.\cite{Petrashov99:3281,Giroud98:R11872,Lawrence96} In contrast to
the calculations presented in this paper, the proximity effect could be
important in weak ferromagnets. Below we will show that most of the recent
experimental results can be explained in terms of the energy-dependence
introduced by interface conductance and the accomplishing change in the spin
accumulation.  It is important to note that these changes are small in
comparison to the total resistance, which is dominated by the long
ferromagnetic wire. Nevertheless they play a dominant role when only the
resistance changes are measured.

The paper is organized in the following way: Section \ref{sec:kin-eq} gives a
description of the diffusive ferromagnetic wire both in the limit of elastic
and inelastic scattering between the electrons. Section \ref{sec:bc} treats
the boundary condition between the ferromagnet and the superconductor which is
crucial to the understanding of the transport properties. The results for the
conductance obtained from the description of the ferromagnetic wire with the
boundary conditions are discussed in Section \ref{sec:results}. Finally we
compare our results with experiments in Section \ref{sec:discussion} and give
our conclusions in Section \ref{sec:concl}.

\begin{figure}
  \begin{center}
    \includegraphics[width=7cm]{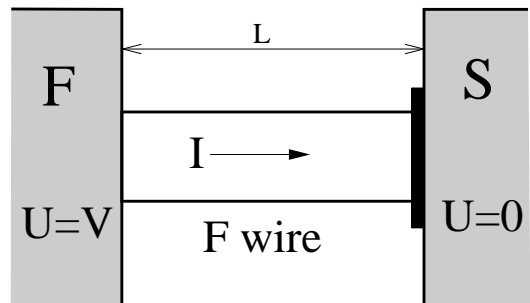}
    \caption{Schematic layout of the mesoscopic ferromagnetic wire. It is
      placed between a ferromagnetic reservoir held at voltage $V$ and
      a grounded superconducting reservoir. The contact to the
      superconductor is an arbitrary connector, characterized by
      spin-dependent conductances in the normal state.}
  \label{fig:fwire}
\end{center}
\end{figure}

\section{Description of the ferromagnet}
\label{sec:kin-eq}

We consider a ferromagnetic diffusive wire connected to an ideal
(ferromagnetic or normal metal) reservoir on one side and to a superconducting
reservoir on the other side as depicted in Fig.~\ref{fig:fwire}. The wire is
characterized by length $L$, cross-section $A$ and spin-dependent
conductivities $\sigma_{\uparrow}$ and $\sigma_{\downarrow}$.  In this Section
we discuss the kinetic equations describing the ferromagnetic wire in the
absence of the proximity effect. We consider collision with impurities to be
the dominant scattering process and use the diffusion approximation. The
electrons in the quasi-one-dimensional wire are described by energy $\epsilon$
and spatial $x$ dependent distribution functions $f_s(\epsilon,x)$ for the two
spin directions $s=+,-$ for spin $\uparrow$ and $\downarrow$, respectively.
The distribution functions obey two coupled Boltzmann equations in the
diffusive limit.  Other scattering mechanisms will be specified in the
following subsections.

Instead of the spin-dependent conductivities, it is convenient to introduce
the total conductivity $\sigma = \sigma_{\uparrow}+\sigma_{\downarrow}$ and
the spin-polarization of the conductivity $\gamma =
(\sigma_{\uparrow}-\sigma_{\downarrow})/\sigma$.  The spin dependent
conductivities are then expressed as $\sigma_s = (1+s\gamma)\sigma/2$.  We
will also make use of the total conductance (resistance) of the ferromagnetic
wire $G_{\text{F}}=A \sigma/L$ ($R_{\text{F}}=1/G_{\text{F}}$).

\subsection{Elastic scattering}
\label{sec:kineq-elastic}

In the elastic scattering case the energy is conserved in the scattering
processes.  This makes it necessary to study the energy dependent distribution
functions in the ferromagnetic wire.  In addition to elastic impurity
scattering we consider only the spin-flip scattering processes, accounted for
by the spin-flip length $l_{\text{sf}}$.  Then, the kinetic equation reads
\begin{equation}
  \frac{d^{2}}{dx^{2}}f_s(\epsilon,x) =
  \frac{1}{l_{\rm sf}^2}\left[f_s(\epsilon,x)-f_{-s}(\epsilon,x)\right]\, .
  \label{kin-eq}
\end{equation}
The current for spin $s$ is given by $(e=|e|)$
\begin{equation}
  \label{eq:f-spin-currents} I_s(x)=-\sigma_s\frac{A}{e}
  \int d \epsilon \frac{df_s(\epsilon,x)}{dx} =\int d\epsilon
  I_s(\epsilon,x)\;.
\end{equation}
This equation defines the spectral current $I_s(\epsilon,x)$.  Electrical and
spin currents are $I_{\text{charge}}=I_{\uparrow}(x)+I_{\downarrow}(x)$ and
$I_{\text{spin}}(x)=I_{\uparrow }(x)-I_{\downarrow}(x)$ and similar for the
spectral currents.  It is convenient to introduce the conductivity-averaged
distribution function
\begin{equation}
  \label{eq:el-distr}
  f_{\text{el}}(\epsilon,x)=
  \frac{\sigma_{\uparrow}}{\sigma}f_{\uparrow}(\epsilon,x)
  +\frac{\sigma_{\downarrow}}{\sigma}f_{\downarrow}(\epsilon,x)
\end{equation}
and the nonequilibrium spin distribution function
\begin{equation}
  \label{eq:spin-distr}
  f_{\text{sp}}(\epsilon,x)=
 f_{\uparrow}(\epsilon,x)-f_{\downarrow}(\epsilon,x)\;.
\end{equation}
The kinetic equations in terms of these functions read
\begin{eqnarray}
  \frac{d^{2}}{dx^{2}}f_{\text{el}}(\epsilon,x) &=&0
  \label{eq:kin-eq1}\\
  \frac{d^{2}}{dx^{2}}f_{\text{sp}}(\epsilon,x) &=&
  \frac{1}{l_{\text{sf}}^{2}}f_{\text{sp}}(\epsilon,x)\;.
  \label{eq:kin-eq2}
\end{eqnarray}
The first equation is the spectral current conservation and the second
describes spatial relaxation of the non-equilibrium spin-distribution.

\subsection{Inelastic scattering}
\label{sec:kineq-inelastic}

The reason to investigate the role of inelastic scattering is the convenient
fact that the ferromagnet is an incoherent metal with rather strong
correlations. Both phonon and electron-electron scattering can mediate
inelastic scattering. In general it is not obvious which should dominate and
both should be treated on equal footing. In order to achieve insight in the
physics it is useful to consider limiting cases as well.

In the limit of strong inelastic scattering we assume that the
electron-electron interaction is stronger than the electron-phonon relaxation.
When a bias voltage is applied, the local electron temperature can therefore
be different from the temperature in the reservoirs.  This transport regime is
relevant when the typical inelastic scattering length is smaller than the
spin-diffusion length.

The electrons relax to a local equilibrium
\begin{equation}
f_{s}(\epsilon ,x)=f(\epsilon ;\mu _{s}(x),T_{\text{el}}(x))\,,
\end{equation}
where $\mu_{s}(x)$ is the spin-dependent chemical potential,
$T_{\text{el}}(x)$ is the local temperature and
\begin{equation}
f(\epsilon ;\mu ,T)=\frac{1}{1+\exp (\left( \epsilon -\mu
\right) /k_{\text{B}}T)}
\label{FD}
\end{equation}
is the Fermi-Dirac distribution function. This makes it possible to integrate
the kinetic equation and the currents over energy and to obtain equations for
the local chemical potentials and electron temperature.

The spin-dependent (electric) current from Eq.~(\ref{eq:f-spin-currents}) is
\begin{eqnarray}
I_{s}(x) & = & - \frac{\sigma}{2e}(1+s\gamma) \frac{d\mu _{s}(x)}{dx} \, .
\end{eqnarray}
Current conservation requires
\begin{equation}
\frac{d^2}{dx^2}\left(\sigma_{\uparrow} \mu_{\uparrow} (x)+
  \sigma_{\downarrow} \mu_{\downarrow} (x)\right) =0 \, .
\label{curcons}
\end{equation}
Spin relaxation occurs within the spin-diffusion length $l_{\text{sf}}$:
\begin{equation}
  \frac{d^2}{dx^2}\left[ \mu _{\uparrow }(x)-\mu _{\downarrow}(x)\right]
  = \frac{1}{l_{\text{sf}}^2}
  \left[ \mu _{\uparrow }(x)-\mu_{\downarrow }(x) \right] \, .
\label{spinrelax}
\end{equation}
The local spin-dependent chemical potentials in the ferromagnet are determined
by (\ref{curcons}) and (\ref{spinrelax}) and the boundary conditions to be
discussed below.

Additionally, we need equations describing energy transport in the system to
account for heating of the electrons. The energy current is
\begin{eqnarray}
  I_{\epsilon}(x) &=&
  -\frac{A}{e^2} \sum_{s} \sigma_{s} \int d\epsilon \epsilon
  \frac{d f_{s}(\epsilon ,x)}{dx} \nonumber \\
&=&\left[ \mu _{\uparrow }(x)I_{\uparrow }(x)+\mu _{\downarrow
}(x)I_{\downarrow }(x)\right] /e+I^Q(x) \, ,
\label{encur}
\end{eqnarray}
where the heat current is
\begin{equation}
I_{\text{Q}}(x)=-\kappa_{\text{Q}} (x) A \frac{dT_{\text{el}}(x)}{dx},
\end{equation}
the heat conductivity $\kappa_{\text{Q}}(x)=\sigma{\cal L}_{0}
T_{\text{el}}(x)$ and the Lorentz number is ${\cal L}_{0}=\frac{\pi^{2}}{3}
\left( \frac{k_{\text{B}}}{e}\right) ^{2}$ .

The conservation law for the energy current dictates
\begin{equation}
  \frac{d}{dx}I_{\epsilon}(x)
  =A\left(\frac{\partial \rho_{\epsilon}(x)}{\partial t}
  \right)_{\text{rel.}}\;,
\label{encons}
\end{equation}
where $\rho_{\epsilon}(x)$ is the local energy density. The energy relaxation
between the electronic system and the phonons at sufficiently low temperatures
is\cite{Allen87:1460}
\begin{equation}
  \left(\frac{\partial\rho_{\epsilon}(x)}{\partial t}\right)_{\text{rel.}}
  =\zeta \left[ \left( k_{\text{B}}T\right) ^{5}-\left(
      k_{\text{B}}T_{\text{el}}(x)\right) ^{5}\right]\; ,
\label{enrel}
\end{equation}
where $\zeta $ parameterizes the strength of the electron-phonon interaction
$\zeta =48\pi \zeta (5)N(\epsilon _{F})\lambda ^{\ast }/ (\hbar ^{3}\omega
_{D}^{2})$, $\zeta (5) \approx 1.04$ is the Riemann zeta function, $N(\epsilon
_{F})$ is the density of states of both spins per unit cell, $\lambda ^{\ast
  }$ is of the same order of magnitude as the electron-phonon coupling
constant $\lambda $, and $\hbar \omega _{D}$ is the Debye energy.

The conservation of energy current (\ref{encons}) together with the expression
for the energy relaxation (\ref{enrel}) give a differential equation for the
local electron temperature which can be solved together with the boundary
conditions to be discussed below.

When the electron-phonon interaction is weak there is no exchange of energy
between the electron and the phonon systems so that the right hand side of
Eq.\ (\ref{encons}) can be set to zero and we have conservation of the energy
current due to the electron transport $dI^{e}(x)/dx=0$. In the opposite limit
of strong electron-phonon interaction the electron temperature equals the
lattice temperature. The differential equation for the energy conservation
with the boundary conditions given above can in these two cases be solved
exactly. In the intermediate regime the equations will be solved numerically.

\section{Boundary conditions}
\label{sec:bc}

The condition that the ferromagnet should be completely incoherent leads to
simplified boundary conditions for the kinetic equations. These boundary
conditions can be derived from the boundary conditions for the quasiclassical
Green's function.\cite{zaitsev:84} A transparent form suitable for diffusive
systems has been presented by Nazarov.\cite{Nazarov99:} We will follow the
spirit and the notation of this paper. A circuit theory for
ferromagnetic-normal metal systems has been presented in Ref.\ 
\onlinecite{Brataas2000:2481}.  A contact is described by a set of
transmission eigenvalues $\{T_n\}$ or, equivalently, by a distribution of the
transmission eigenvalues $\rho(T)$. The boundary condition at the contact is
expressed through a conservation law for the matrix current in the Keldysh
formulation. In the framework of superconductivity it is a $4\times 4$-matrix
comprising $2\times 2$ Keldysh space and $2 \times 2$ particle-hole (Nambu)
space. The matrix current through the FS-contact is\cite{Nazarov99:}
\begin{equation}
  \label{eq:matrix-current}
  \check{I}=-\frac{2e}{\pi\hbar}\sum\limits_n T_n
  \frac{\left(\check{G}_{\text F}\check{G}_{\text{S}}
      -\check{G}_{\text S}\check{G}_{\text{F}}\right)}{
    4+T_n(\check{G}_{\text F}\check{G}_{\text{S}}
    +\check{G}_{\text S}\check{G}_{\text{F}}-2)}
  \;.
\end{equation}
This matrix current has to be equated to the diffusive matrix current entering
the contact from either side.  The two sides of the contact are characterized
by the Keldysh matrix Green's functions $\check{G}_{\text S}$ and
$\check{G}_{\text F}$, which we will specify to be the superconducting
reservoir and the ferromagnetic wire, respectively.  The Keldysh-Nambu matrix
Green`s function of the superconductor in equilibrium is

\begin{equation}
  \label{eq:s-greenfunc}
  \check{G}_{\text{S}}(\epsilon)=\left(\begin{array}{cc}\hat{G}^{\text{R}}_{\text{S}}(\epsilon)
      &\hat G^{\text{K}}_{\text{S}}(\epsilon) 
      \\ 0 
      & \hat G^{\text{A}}_{\text{S}}(\epsilon)  \end{array} \right) \, .
\end{equation}
A similar structure holds for any matrix in Keldysh space. In local
equilibrium the Keldysh (1,2) component in Nambu space is
\begin{equation}
  \hat G^{\text{K}}_{\text{S}}(\epsilon) = (\hat
  G^{\text{R}}_{\text{S}}(\epsilon)- \hat
  G^{\text{A}}_{\text{S}}(\epsilon))
  \left(1-2f^{\text{S}}(\epsilon)\right) \, ,
\end{equation}
where $f^{\text{S}}(\epsilon)=[1+\exp(\epsilon/k_{\text{B}}T)]^{-1}$
is the quasi-particle distribution function in the superconductor and
we have set the chemical potential in the superconductor to zero.
$\hat G^{\text{R}}_{\text{S}}(\epsilon)$ and $\hat
G^{\text{A}}_{\text{S}}(\epsilon)$ are retarded and advanced Nambu
Green's functions determining the spectral properties of the
superconductor.  In the BCS case with a real order parameter they are
\begin{eqnarray}
  \label{eq:grga-bcs} \hat G^{\text{R}}(\epsilon)=
  -\left(\hat{G}^{\text{A}}(\epsilon)\right)^*=
  \frac{(\epsilon+i0)\hat\tau_3-i\Delta\hat\tau_1}{
    \sqrt{(\epsilon+i0)^2-\Delta^2}}\;.
\end{eqnarray}
The diagonal component represents the normal retarded Green's function
whereas
the off-diagonal component is conventionally called the anomalous Green's
function.  On the ferromagnetic side we completely neglect the proximity
effect leading to the spectral functions $\hat
G^{\text{R}}_{\text{F}}=\hat\tau_3=-\hat G^{\text{A}}_{\text{F}}$. The
absence
of an anomalous component is a result of the absence of the proximity
effect.
The Keldysh component accounts for the spin-dependent non-equilibrium
distribution:
\begin{equation}
  \label{eq:f-greenfunc}
  \hat G^{\text{K}}_{\text{F}}(\epsilon)=2\left(
    \begin{array}[c]{cc}
      1-2f^{\text{F}}_\uparrow(\epsilon) &0\\
      0 & 1-2f^{\text{F}}_\downarrow(-\epsilon)
    \end{array}\right)\;,
\end{equation}
where $f^{\text{F}}_{\uparrow}(\epsilon)$ and
$f^{\text{F}}_{\downarrow}(\epsilon)$ are the quasi-particle distribution
functions close to the interface on the ferromagnetic side.  The spectral
electrical current is determined by the Keldysh-component of the matrix
current according to
\begin{equation}
  \label{eq:current1}
  I_{\text{el}}(\epsilon)=\frac{1}{4e}\text{Tr}\left[
  \hat{\tau}_{3}\hat{I}^{\text{K}}(\epsilon)\right]\;.
\label{eq:speccur}
\end{equation}
Eq.\ (\ref{eq:f-greenfunc}) and Eq.\ (\ref{eq:speccur}) suggest a
representation of the diagonal components of the Keldysh component of the
matrix current in the form
\begin{equation}
  \label{eq:current2}
  \hat{I}^{\text{K}}(\epsilon)=\left(
    \begin{array}[c]{cc}
      I_\uparrow(\epsilon)&\cdots\\
      \cdots&I_\downarrow(-\epsilon)
    \end{array}\right)\;.
\end{equation}
Now we are in the position to calculate the spin-resolved currents through the
contact. Performing the calculations along the lines of Ref.
\onlinecite{Nazarov99:} we find the spectral spin-dependent current
\begin{eqnarray}
  \label{eq:spin-currents} I_{s}(\epsilon)&=&
  \frac{G_{\text{QP}}(\epsilon)}{2e}
  \left(f^S(\epsilon)-f_{s}^{F}(\epsilon)\right)\\&&\nonumber
  +\frac{G_{\text{A}}(\epsilon)}{4e}\left(1-f_{s}^{F}(\epsilon)
  -f_{-s}^{F}(-\epsilon)\right)\;.
\end{eqnarray}
The quasiparticle conductance $G_{\text{QP}}(\epsilon)$ and the Andreev
conductance $G_A(\epsilon)$ are determined by the properties of the contact
and the spectral properties of the two metals connected by the contact. The
distribution of transmission eigenvalues can be incorporated in a single
characteristic complex function
\begin{eqnarray}
  \label{eq:z-of-e}
  Z(x)=\frac{ e^2}{\pi\hbar}\sum\limits_n
  \frac{T_n}{2+T_n(x-1)},
\end{eqnarray}
where $x(\epsilon)=\text{Tr}\{\hat G^{\text{R}}_{\text{S}}(\epsilon),\hat
G^{\text{R}}_{\text{F}}(\epsilon))\}/4$. The conductances are
\begin{eqnarray}
  \label{eq:gqp-def}
  G_{\text{QP}}(\epsilon)&=&\text{Re}Z(x)\text{Re}x+
  \frac{\text{Im}Z(x)}{\text{Im}x}\text{Im}^2\sqrt{1-x^2} \, ,\\
  \label{eq:ga-def}
  G_{\text{A}}(\epsilon) &=&
  -\frac{\text{Im}Z(x)}{\text{Im}x}\left|1-x^2\right| \, .
\end{eqnarray}

\begin{figure}[htbp]
  \begin{center}
    \includegraphics[width=7cm]{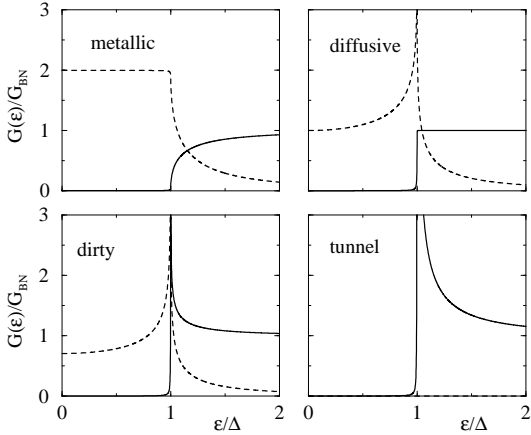}
    \caption[]{Spectral conductances for different types of contacts. The
    solid curves denote the quasiparticle conductance
    $G_{\text{QP}}(\epsilon)$ and the dashed curves the Andreev conductance
    $G_{\text{A}}(\epsilon)$. The contact types as indicated in the figure are
    metallic junction (all $T_n=1$), diffusive contact (transmission
    distribution as defined in Eq.~(\ref{eq:diffusive})), dirty interface
    (Eq.~(\ref{eq:dirty})), and tunnel junction (all $T_n\ll 1$).}
    \label{fig:spectral}
  \end{center}
\end{figure}

The contact is characterized by a transmission distribution, which leads to
contact-specific energy dependences of the conductances. The normal state
conductance is $G_{\text{BN}}=(e^2/2\pi \hbar) \sum_n T_n$. For a ballistic
model contact all transmission eigenvalues are equal to one for the
propagating channels and zero otherwise and $\sum_n T_n=N$, where $N$ is the
number of propagating channels. The distribution function in the case of a
dirty interface is\cite{Schep97:3015}
\begin{equation}
  \label{eq:dirty}
  \rho(T) = \frac{\hbar}{e^2} G_{\text{BN}} \frac{1}{T^{3/2}\sqrt{1-T}}
\end{equation}
and in the case of a diffusive contact the distribution
is\cite{Dorokhov82:259}
\begin{equation}
  \label{eq:diffusive}
  \rho(T) = \frac{\hbar}{2e^2} G_{\text{BN}} \frac{1}{T \sqrt{1-T}} \, .
\end{equation}
Finally, for a tunnel conductance a perturbation expansion in terms of the
small transmission eigenvalues can be performed. We list the characteristic
function $Z(x)$ for a number of generic contacts in Table~(\ref{tab:1}):
Tunnel junction, ballistic contact, diffusive contact and dirty interface.  In
the case of an incoherent metal on one side (i.e. $\hat
G^{\text{R}}_{\text{F}}=\hat\tau_3$), the argument of the characteristic
function reduces to $x=\text{Tr}\hat\tau_3\hat{G}_{\text{S}}^{\text{R}}/2$.
The result in this case is demonstrated explicitely in Table~\ref{tab:1} for a
contact of a BCS-superconductor with spectral functions (\ref{eq:grga-bcs}).
The energy dependence of these spectral conductances is depicted in
Fig~\ref{fig:spectral}. Below the superconducting gap only the Andreev
conductance is nonzero, gradually decreasing from the value of
$2G_{\text{BN}}$ for the metallic junction to zero in the tunnel junction.
Above the gap the Andreev conductance vanishes rather quickly $\sim
1/\epsilon^2$. Also quasiparticle transport becomes possible and, thus,
spin-transport into the superconductor.

\begin{figure}
  \begin{center}
    \includegraphics[width=7cm]{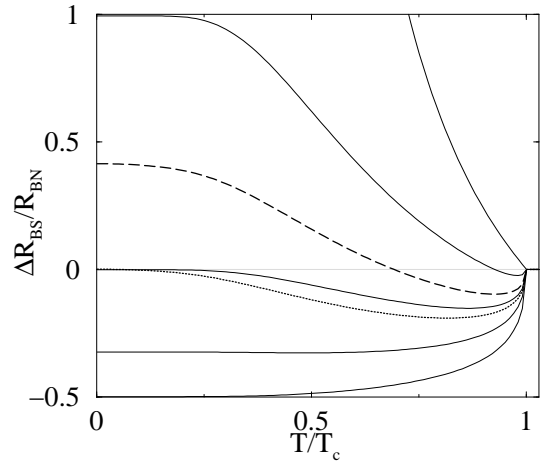}
    \caption[]{Resistance change of different types of contacts between a
      normal metal reservoir and a superconducting reservoir. The mixed
      contact (solid) varies from ballistic ($q=0$) to tunnel ($q=1$)
      from bottom to top. Intermediate values are ($q= 0.25, 0.5,
      0.75$).  The diffusive contact is shown by the short-dashed line
      and the dirty contact by a long-dashed line (see Table \ref{tab:1}
      for a definition of these contacts).}
    \label{fig:cond-t-dep}
  \end{center}
\end{figure}

The properties of these contacts are demonstrated by the temperature
dependence of the linear conductance following from
\begin{equation}
  \label{eq:cond-temp-dep}
  G_{\text{BS}}(T)=\int \! \! d\epsilon (G_{\text{QP}}(\epsilon)+
  G_{\text{A}}(\epsilon))
  \left(-\frac{\partial f(\epsilon,0,T)}{\partial \epsilon}\right)\;.
\label{eq:BScon}
\end{equation}
This is the conductance that would be measured if the contact would be placed
between a normal reservoir and a superconducting reservoir.  The temperature
dependence of the contact conductance (\ref{eq:BScon}) is shown in
Fig.~\ref{fig:cond-t-dep}. The dashed and dotted lines show the conductance of
the diffusive contact and the dirty interface, respectively.  The resistance
of the diffusive contact shows the well known reentrant behavior, {\em i.e.}
it reaches the normal state conductance at zero temperature.\cite{AVZ:79} The
resistance of the dirty interface after a small drop below the critical
temperature is higher than the normal state value and saturates at low
temperature at $\sqrt{2} R_{\text{BN}}$.\cite{Schep97:3015}

\end{multicols}
  \begin{table}[htbp]
    \begin{center}
      \begin{tabular}[t]{|c|c||c||c|c|c|}
        Contact &
        $\rho(T)/G_{\text{N}}$ &
        $Z(\epsilon)/G_{\text{N}}$ &
        \multicolumn{3}{c|}{F-S-Contact}\\\hline
        &&&
        $\frac{G_{\text{A}}}{G_{\text{N}}}$ $(\epsilon<\Delta)$ &
        $\frac{G_{\text{A}}+G_{\text{QP}}}{G_{\text{N}}}$ $(\epsilon>\Delta)$ &
        $\frac{G_{\text{QP}}}{G_{\text{N}}}$ $(\epsilon>\Delta)$
        %% tunnel
        \\\hline\hline tunnel & $\displaystyle T_n \ll 1$ &
        $\displaystyle 1$ & $0$ &
        $\displaystyle \frac{\epsilon}{\xi}$ &
        $\displaystyle \frac{\epsilon}{\xi}$
        %% ballistic
        \\\hline\hline
        ballistic &
        $\displaystyle T_n=1$ &
        $\displaystyle \frac{2}{1+x}$ &
        $2$ & $\displaystyle \frac{2\epsilon}{\epsilon+\xi}$ &
        $\displaystyle \frac{2\xi}{\epsilon+\xi}$
        %% diffusive
        \\\hline\hline
        diffusive &
        $\displaystyle \frac{\hbar \pi}{2e^2}\frac{1}{T\sqrt{1-T}}$ &
        $\displaystyle \frac{\arccos(x)}{\sqrt{1-x^2}}$ &
        $\displaystyle \frac{\Delta}{2\epsilon}
        \ln\left(\frac{\Delta+\epsilon}{\Delta-\epsilon}\right)$ &
        $\displaystyle \frac{\epsilon}{2\Delta}
        \ln\left(\frac{\epsilon+\Delta}{\epsilon-\Delta}\right)$ &
        $1$
        %% dirty interface
        \\\hline\hline
        dirty interface &
        $\displaystyle\frac{\hbar}{e^2} \frac{1}{T^{3/2}\sqrt{1-T}}$ &
        $\displaystyle\sqrt{\frac{2}{1+x}}$ &
        $\displaystyle \frac{\Delta}{\sqrt{\xi(\xi+\Delta)}}$ &
        $\displaystyle\sqrt{\frac{\epsilon+\xi}{2\xi}}$ &
        $\displaystyle\frac{\sqrt{2}\epsilon}{\sqrt{\xi(\epsilon+\xi)}}$
        %% abbrevs
        \\\hline\hline abbreviations & & $x=\frac14
        \text{Tr}\{\hat G^{\text{R}}_{\text{S}},\hat
        G^{\text{R}}_{\text{F}}\}$ &
        \multicolumn{2}{c}{$\xi=\sqrt{\left|\epsilon^2-\Delta^2\right|}$}&
      \end{tabular}
      \caption{Spectral conductances of different
        generic contacts defined by the transmission distributions in the
        second column. The characteristic function $Z$ of the contact was
        defined in Eq.\ (\ref{eq:z-of-e}). As an example we present
electrical
        and spin conductance for a contact between a ferromagnetic metal and
a
        BCS Superconductor in the last two columns. For energies below the
        superconducting gap $\Delta$ the quasiparticle conductance $G_{QP}$
        vanishes for all contacts. Note, that the energy argument of all
        quantities $\epsilon$ is understood to be the absolute value of the
        energy.}
      \label{tab:1}
    \end{center}
  \end{table}
\begin{multicols}{2}

Additionally, we introduce mixed contacts as a model for an inhomogeneous
interface with distributed regions with low and high transparency.  The
relative admixture $q$ of a tunnel and $(1-q)$ of a ballistic contact allows
switching continuously from one limit to the other, covering approximately the
universal cases of a diffusive contact ($q\approx 0.5$) and a dirty interface
($q\approx 1/\sqrt{2}$ with a single parameter $q$. A common feature of the
temperature dependence of all these contacts except the tunnel junction is
that right below $T_{\text{c}}$ the resistance drops. At lower temperatures
the resistance increases again except in the case of the purely ballistic
contact.  The drop of resistance of these contacts close to $T_{\text{c}}$ can
be traced back to the temperature dependence of the superconducting order
parameter $\Delta(T)$.  The resistance drop is caused by the leading order
contribution of the change in the superconducting gap $ \Delta(T) \propto
(1-T/T_{\text{c}})^{1/2}$ to the Andreev contribution and the
conductance.\cite{delta}

The boundary condition presented so far imply that the transmission ensembles
and the number of channels are the same for the two spin species. In reality
the transmission matrices for spin-up and spin-down states can be different.
A microscopic calculation of the transmission eigenvalues is beyond the scope
of the present paper.  We will therefore heuristically generalize the boundary
conditions to spin-dependent interfaces by taking different transmission
ensembles for the two spin directions. These ensembles can differ in the total
number of channels and/or in the transmission distribution. Thus, we replace
the spin-dependent current through the interface (\ref{eq:spin-currents}) by
\begin{eqnarray}
  I_{\text{s}}(\epsilon) & = & \frac{G_{\text{s}}(\epsilon)}{2e}
  (f^{S}(\epsilon)-f^{\text{F}}_{\text{s}}(\epsilon,0))
  \\\nonumber &&
  \frac{G_{\text{A}}(\epsilon)}{4e}
  \left(1-f^{\text{F}}_{\text{s}}(\epsilon,0)
  +f^{\text{F}}_{\text{-s}}(-\epsilon,0)\right)\;,
  \label{eq:fs-spin-currents}
\end{eqnarray}
In general the spin-dependent quasi-particle conductances
$G_{\uparrow}(\epsilon)$ and $G_{\downarrow}(\epsilon)$ entering the
first term are of different magnitude and have different energy
dependences.  Similar as for the ferromagnetic wire we introduce the
total conductance of the boundary $G_B(\epsilon) =
G_{\uparrow}(\epsilon)+G_{\downarrow}(\epsilon)$ and a dimensionless
factor $\gamma_B(epsilon) = (G_{\uparrow}(\epsilon) -
G_{\downarrow}(\epsilon)) / G_B(\epsilon)$, which we call polarization
of the boundary conductance. Since the definitions (\ref{eq:gqp-def})
of quasi-particle and (\ref{eq:ga-def}) of Andreev conductance have
been derived for a spin-degenerate interface these definition are not
valid anymore for spin-dependent interface scattering. It is, however,
reasonable to assume that the energy dependence of all conductances is
well approximated by the {\em same} transmission ensemble, but {\em
  different} numbers of channels. We can motivate this choice by the
fact that in the experiments that we have in mind the interfaces are
strongly disordered regions, with a possible formation of an alloy
layer extending over several monolayers. In such contacts the number
of channels is more or less controlled by the differences of the cross
sections of the Fermi surface. But, on the other hand, the
transmission ensemble and, hence, the energy dependence of the
conductance is not expected to vary much in typical disordered
contacts on the scale of the superconductor gap.

We will in the following only take into account the differences in magnitude,
but not in energy dependence.  In the language of transmission distributions
this means that the distributions are the same, but the number of channels
differ. In this approximation the spin-polarization of the boundary
conductance $\gamma_{\text{B}}$ is energy independent. The energy dependence
of the Andreev conductance follows from the same transmission ensemble, but
its magnitude will be reduced in comparison to the unpolarized case.  It is
important to notice that the boundary polarization and the polarization of the
ferromagnetic wire need not to have the same sign, since they are
parametrically independent. The possibility of this is demonstrated by
microscopic numerical calculations.\cite{schep}

\section{Results and discussions}
\label{sec:results}

We solve the kinetic equations presented in Sec.~\ref{sec:kin-eq} in
the three cases:
\begin{itemize}
\item[A.] purely elastic scattering
\item[B.] inelastic scattering in linear response
\item[C.] inelastic scattering in nonlinear response
\end{itemize}
The boundary condition on the superconducting side of the wire has been
derived in Sec.~\ref{sec:bc}. The boundary conditions at the ferromagnetic
reservoir are
\begin{equation}
  \label{eq:res-bc}
  f_{\uparrow}(\epsilon,-L)=f_{\downarrow}(\epsilon,-L)
  =f(\epsilon;eV,T)\;,
\label{eq:boundfer}
\end{equation}
where $f(\epsilon;eV,T) =(\exp((\epsilon-eV)/k_{\text{B}}T)+1)^{-1}$ is the
Fermi-Dirac equilibrium distribution at a constant voltage $V$ and
temperature $T$.

In the case of inelastic scattering Eq.~(\ref{eq:boundfer}) also implies that
the electron temperature equals the lattice temperature in the ferromagnetic
reservoir $T_{\text{el}}(x=-L)=T $.  The other boundary condition for the
electron temperature comes from the conservation of energy current in the
ferromagnet and into the superconductor.

As a reference we calculate the resistance of the system in the normal state
\begin{equation}
  R_{\text{FN}} =R_{\text{F}} + R_{\text{BN}} + R_{\text{sf}}
  \frac{(\gamma-\gamma_{\text{B}})^2}{1+R_{\text{sf}}/R_{\text{BN}}}
  \, ,
\label{RFN}
\end{equation}
The third term is due to the spin-accumulation in the ferromagnetic wire
determined by the `spin-flip resistance' $R_{\text{sf}} =
1/G_{\text{sf}}=R_{\text{F}} l_{\text{sf}}/L \tanh(l_{\text{sf}}/L)
(1-\gamma^2)$. In the limit of a weak ferromagnet $\gamma^2\ll 1$ and a short
spin-flip relaxation length $l_{\text{sf}}\ll L$ the spin-flip resistance
reduces to $R_{\text{sf}} \approx R_{\text{F}} l_{\text{sf}}/L$, {\em i.e.}
the resistance of a piece of the ferromagnetic wire of length $l_{\text{sf}}$.
We see that the excess resistance due to the spin-accumulation increases with
increasing asymmetry between the polarization of the bulk conductivity and the
polarization of the interface conductance.  The expression (\ref{RFN}) will be
used in the following to calculate resistance changes below the transition to
the superconducting state:
\begin{equation}
  \label{eq:res-change}
  \Delta R_{\text{FS}}(T,V)=R_{\text{FS}}(T,V)-R_{\text{FN}}\;.
\label{diffres}
\end{equation}
The differential resistance is defined by
\begin{equation}
  \label{eq:diff-res}
  R_{\text{FS}} (T,V)=\left( \frac{\partial I(T,V)}{\partial V}\right) ^{-1}
\;.
\end{equation}
In the linear response regime we will omit the arguments of the differential
resistance $R_{\text{FS}}\equiv R_{\text{FS}}(T,V\to 0)$.

In the following analysis it will be useful to define the following
temperature dependent average
\begin{eqnarray}
  \label{eq:average}
  \langle\cdots\rangle = \int_{-\infty}^{\infty}  \cdots
  \left(-\frac{\partial f(\epsilon;0,T)}{\partial \epsilon}\right)d\epsilon
\label{tempavg}
\end{eqnarray}
This average occures, {\em e.g.}, in the temperature dependent conductances of
a contact between an incoherent metal and a superconductor
(\ref{eq:cond-temp-dep}).

\subsection{Elastic scattering}

When the scattering in the wire is elastic, the general solution of
(\ref{eq:kin-eq1}) satisfying (\ref{eq:res-bc}) may be written as
\begin{eqnarray}
  \label{eq:el-sol1}
  \nonumber
  f_{\text{el}}(\epsilon,x)&=&
  \frac{G_{\text{FS}}(\epsilon)}{G_{\text{F}}}
  \left(f(\epsilon;0,T)-f(\epsilon;eV,T)\right)
  \left(1+\frac{x}{L}\right)
  \\&&
  +f(\epsilon;eV,T).
\end{eqnarray}
The spatially independent spectral conductance $G_{\text{FS}}(\epsilon)$
determines the current through the structures and remains to be found.  The
solution of the second kinetic equation (\ref{eq:kin-eq2}) satisfying the
boundary condition (\ref{eq:res-bc}) is
\begin{equation}
f_{\text{sp}}(\epsilon,x)=2\alpha \sinh\left(\frac{L+x}{l_{s}}\right) \;,
\end{equation}
where the parameter $\alpha$ should be found from the continuity of the spin
currents into the superconductor (\ref{eq:fs-spin-currents}) and ferromagnetic
wire (\ref{eq:f-spin-currents}). We find the electrical current
\begin{eqnarray}
  I(T,V) & = & \frac{1}{2e} \int d\epsilon G_{\text{FS}}(\epsilon)
\times \nonumber \\ && \left[
1-f(\epsilon;eV,T)-f(-\epsilon;eV,T)\right] \, .
\label{eq:tot-current}
\end{eqnarray}
This expression shows that the spectral conductance determines the transport
in each energy slice depending on the difference in occupation of states at
this energy in the reservoirs. This form is analogous to the classical
definition of a conductance as the proportionality factor between current and
voltage difference.

The spectral conductance is given by
\begin{eqnarray}
  \label{eq:gfs}
  \frac{1}{G_{\text{FS}}(\epsilon)} & = &
  \frac{1}{G_{\text F}}
  +\frac{1}{G_{\text{QP}}(\epsilon)+G_{\text{A}}(\epsilon)}
  \\\nonumber&&
  +\frac{\left(\gamma-\gamma_{\text{B}}
      \frac{G_{\text{QP}}(\epsilon)}{
        G_{\text{QP}}(\epsilon)+G_{\text{A}}(\epsilon)}\right)^{2}}{ 
    G_{\text{sf}}
    +G_{\text{QP}}(\epsilon)\left(1-\gamma_{\text{B}}^2
      \frac{G_{\text{QP}}(\epsilon)}{
        G_{\text{QP}}(\epsilon)+G_{\text{A}}(\epsilon)}\right)}\;.
\end{eqnarray}
In the general case the full expression has to be used to calculate the
resistance change in the superconducting state.

When the ferromagnetic wire dominates the resistance of the whole structure a
simplified expression for the linear resistance change may be obtained. We
first limit the discussion to the case of a weak ferromagnet and vanishing
boundary polarization to obtain
\begin{eqnarray}
  \label{eq:res-change1}
  \Delta R_{\text{FS}}& =& 
  \left\langle 
    \frac{1}{G_{\text{QP}}(\epsilon)+G_{\text{A}}(\epsilon)}
  \right\rangle
    -R_{\text{BN}}\\\nonumber&&
  +\gamma^2
  \left[
    \left\langle 
      \frac{1}{G_{\text{sf}}+G_{\text{QP}}(\epsilon)}
    \right\rangle
    -\frac{1}{G_{\text{sf}}+G_{\text{BN}}}
  \right]  \;.
\end{eqnarray}
We see that the resistance change consists of two contributions. The first is
the resistance change due to the change of the boundary resistance, which
would also be present in the absence of spin polarization. Note, however, that
this term can be qualitatively different from the case of a normal metal wire
in contact to a superconductor since in this case the proximity effect would
not be negligible. The second term accounts for the difference in spin
accumulation between normal and superconducting state.

\begin{figure}
  \begin{center}
    \includegraphics[width=7cm]{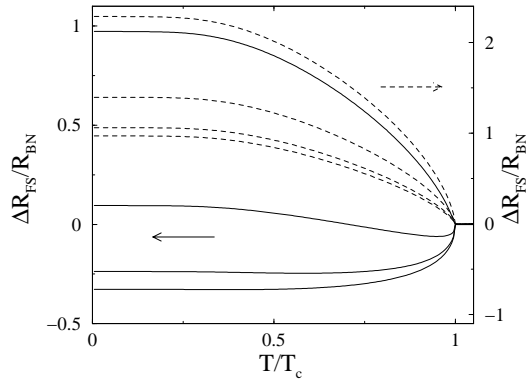}
    \caption{Temperature dependence of the resistance change
      of the F-wire attached to a superconducting reservoir.  Results
      are presented for two mixed contacts with $q=0.5$ (solid lines)
      and $q=0.25$ (dashed lines). The conductivity polarization
      $\gamma$ of the wire is changed from $0$ to $0.6$ in steps of
      $0.2$ from the bottommost curve to the topmost curve for both
      contacts. Other parameters are $l_{\text{sf}}=0.03L$, and
      $R_{\text F}=100 R_{\text{BN}}$. Clearly spin accumulation leads
      to an enhanced resistance in both cases. The resistance of the
      $q$=0.25-contact is more or less uniformly increased.  The effect
      of spin accumulation is much more dramatic for the
      $q$=0.75-contact. The monotonic resistance decreases in the
      unpolarized case is first turned into an reentrant behavior for
      small polarization overshooting the normal state resistance
      slightly at low temperatures.  Increasing $\gamma$ further leads
      to an increased resistance for all temperature. Note that this
      behavior resembles that of a $q$=0.25-contact, if properly
      rescaled.}
    \label{fig:el-spinaccu}
  \end{center}
\end{figure}

First we discuss the influence of spin accumulation on the FS-resistance for a
spin-degenerate interface. In Fig.~\ref{fig:el-spinaccu} resistance changes
for two types of contacts are shown for different polarizations of the
ferromagnet. Solid curves are for a relatively good contact ($q=0.75$) and
dashed curves for a less transparent contact ($q=0.25$). In this plot the
total resistance of the system is dominated by the resistance of the
ferromagnetic wire $R_{\text F}=100R_{\text{BN}}$ and the spin-relaxation
length is $l_{\text{sf}}=0.03L$, resulting in a spin accumulation resistance
$R_{\text{sf}}\approx 3R_{\text{BN}}$. Accordingly, the resistance change is
normalized to $R_{\text{BN}}$ to show the relevant scale of the effect
produced by the superconducting transition.  For both contacts spin
accumulation (increasing from the bottom to the top curves) leads to an
enhancement of the resistance. Specifically the low temperature resistance is
well accounted for by Eq.~(\ref{eq:res-change1}) in the limit $T\to 0$:
\begin{equation}
  \label{eq:rfs-zerotemp}
  \Delta R_{\text{FS}}(T=0) = \frac{1}{G_{\text{A}}(0)}-R_{\text{BN}}
  +\gamma^2\frac{R_{\text{sf}}}{1+R_{\text{BN}}/R_{\text{sf}}}
  \;.
\end{equation}
The second term of this equation shows that the spin-accumulation always
enhances the resistance, maximally by an amount $\gamma^2 R_{\text{sf}}$.
The enhancement for the $q=0.25$-contact has a uniform temperature dependence
and does not change qualitatively. This is different for the $q=0.75$ contact.
Here the resistance decreases monotonically in the unpolarized case as a
result of the Andreev enhanced conductance. A small polarization
$\gamma\approx 0.2-0.4$ results in a nonmonotonic temperature dependence, {\em
  i.e.} an increase of resistance at lower temperatures. This can lead (for
specific parameters) to a reentrant behavior of the resistance change, even
overshooting the normal state value for larger spin-accumulation. At even
higher spin polarizations $\gamma^2 \gtrsim R_{\text{BN}}/R_{\text{sf}}$ the
Andreev contribution is completely masked and the resistance increases
monotonically. This behavior resembles that of a less transparent contact if
the absolute scale is properly chosen.

\begin{figure}
  \begin{center}
    \includegraphics[width=7cm]{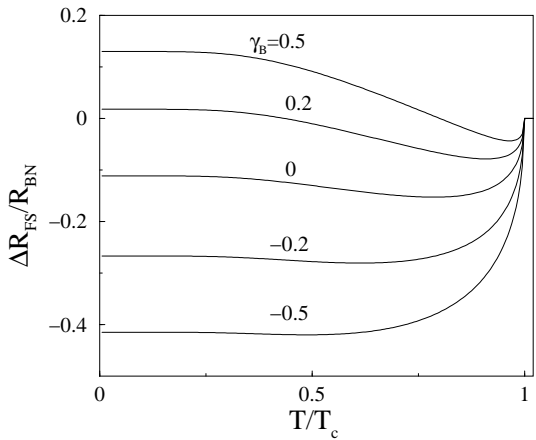}
    \caption{Effect of the relative polarizations on the resistance change.
      The contact is a mixed contact with $q=0.75$, other parameters are
      $\gamma=0.3$, $l_{\text{sf}}=0.03L$, and $R_{\text F}=100 R_{\text{BN}}$.
      The polarization of the boundary conductance $\gamma_{\text{B}}$ is
      varied between a symmetric configuration $\gamma_{\text{B}}=0.5$ and an
      antisymmetric configuration $\gamma_{\text{B}}=-0.5$. At the same time
      the Andreev-conductance is rescaled by a factor $1-\gamma_B^2$ to
      account for the smaller number and transmission of Andreev channels.
      For large antisymmetric polarization the resistance decrease exceeds the
      decrease of the corresponding normal metal-superconductor contact.}
    \label{fig:el-poldep}
  \end{center}
\end{figure}

Let us now discuss the effect of the interface polarization
$\gamma_{\text{B}}$ on the resistance change. In Fig.~\ref{fig:el-poldep} the
temperature dependent resistance of a $q=0.75$ contact is shown for different
interface polarizations. Other parameters are $R_{\text F}=100R_{\text{BN}}$,
$l_{\text{sf}}=0.03L$, and $\gamma=0.3$. The interface polarization
$\gamma_{\text{B}}$ changes from the symmetric value $+0.5$ to the
antisymmetric value $-0.5$, as indicated in the plot. The reduction of the
Andreev conductance by the spin-dependent interface resistance is taken into
account by a phenomenological renormalization factor
$(1-\gamma_{\text{B}}^2)$. To gain some insight it is useful to look at the
low temperature limit of the resistance change in the limit $R_{\text{BN}}\ll
R_{\text{sf}}$. From Eq.~(\ref{eq:gfs}) it follows that
\begin{equation}
  \label{eq:rfs-pol-zerotemp}
  \Delta R_{\text{FS}}(T=0) = \frac{1}{G_{\text{A}}(0)}-R_{\text{BN}}
  +R_{\text{sf}}(4\gamma\gamma_{\text{B}}-\gamma_{\text{B}}^2)\;.
\end{equation}

The spin-dependent contribution depends on the relative sign of the two
polarizations $\gamma$ and $\gamma_{\text{B}}$ and can also be negative (if
$4\gamma\gamma_{\text{B}}<\gamma_{\text{B}}^2$). This effect is seen from the
lower two curves in Fig.~\ref{fig:el-poldep} with an antisymmetric interface
polarization. An increasing interface polarization leads to a lowering of the
resistance change, despite the increase of the resistance due to the
renormalization of the Andreev conductance. It is worthwhile noting that for
the largest negative interface polarization shown ($\gamma_{\text{B}}=-0.5$)
the total resistance drop is {\em larger} than the resistance drop which would
result from the pure Andreev reflection in the absence of spin polarization of
the interface and the F-wire. This apparent contradiction to the intuition
that any spin-accumulation should decrease the Andreev-caused resistance drop
stems from the fact that we plot the resistance {\em change} below the
superconducting transition. The contradiction is resolved by noting that the
{\em total} resistance $R_{\text{FS}}(T)=R_{\text{FN}}+\Delta
R_{\text{FS}}(T)$ is always higher than for the unpolarized case. However, in
a real experiment (with fixed polarizations) the Andreev conductance in the
absence of a polarization can not be measured separately. It may therefore
appear that the measured resistance drop is larger than one would expect from
a simple estimate of the reduction of the interface resistance due to Andreev
reflection.

\subsection{Inelastic scattering - linear response}

We will now proceed to study the case of inelastic scattering in the
ferromagnetic wire. It is assumed that the current in the ferromagnet is
weakly polarized, $\gamma \ll 1$. In order to simplify the discussions we
disregard the possible asymmetry in the interface transparency in the
following discussions and set $\gamma_{\text{B}}=0$. An extension is
straightforward.

An analytical expression for the total conductance of the system can be found
in the linear response regime. In this regime the effects of electron heating
vanish since they will only contribute to the current in higher orders of the
source-drain bias. The coupled equations for the spin-dependent chemical
potential distributions and the electron temperature are simplified by letting
$T_{\text{el}}(x) \rightarrow T$. By solving (\ref{curcons}) and
(\ref{spinrelax}) together with the boundary condition
(\ref{eq:spin-currents}) and (\ref{eq:boundfer}) we find the linear response
resistance.  Assuming a weak ferromagnet, $\gamma^2 \ll 1$, and a small
interface resistance compared to the resistance of the ferromagnetic wire
$R_{\text{BN}} \ll R_{\text{F}}$ the resistance change can be written as
\begin{eqnarray}
  \label{RFSlinINEL}
  \Delta R_{\text{FS}}(T) & = &
  \frac{1}{\left\langle
      G_{\text{QP}}(\epsilon)+G_{\text{A}}(\epsilon)\right\rangle}
  -R_{\text{BN}} \\ \nonumber &&
  +\gamma^2\left[
    \frac{1}{G_{\text{sf}}+\left\langle
G_{\text{QP}}(\epsilon)\right\rangle}
    -\frac{1}{G_{\text{sf}}+G_{\text{BN}}}\right]\,.
\end{eqnarray}
The first two terms in (\ref{RFSlinINEL}) are due to the effective interface
resistance between the ferromagnet and the superconductor and the third term
is due to the spin-accumulation. The latter term vanishes when $\gamma
\rightarrow 0$ or $l_{\text{sf}}/L \rightarrow 0$.  This equation has to
compared with Eq.~(\ref{eq:res-change}) for the case of purely elastic
scattering.  Only the quasi-particle conductance enters the spin-accumulation
contribution since spins cannot be injected into the superconductor by means
of the Andreev process.  The temperature averaged conductances directly
determine the temperature dependence of the total resistance in the case of
dominant inelastic scattering processes.  The quasi particle conductance
vanishes at zero temperature since then no spin-current can propagate into the
superconductor. At zero temperature $\langle
G_{\text{QP}}(\epsilon)+G_{\text{A}}(\epsilon)\rangle = G_{\text{A}}(0)$ and
the resistance the FS system in the case of inelastic scattering
(\ref{RFSlinINEL}) equals the result in the case of elastic scattering
(\ref{eq:rfs-zerotemp}).

The results with inelastic scattering in general differ from those with purely
elastic scattering when the temperature is non-zero or when the current is
measured in the non-linear source-drain response regime. The remarkable
difference between Eq.~(\ref{eq:res-change}) and Eq.~(\ref{RFSlinINEL}) is the
way the thermal averaging is carried out. {\em E.g.}  in the first term we
have to average the inverse contact conductance in the case of elastic
scattering, whereas we {\em first} have to average the conductance and than
invert the result in the case of inelastic scattering.  A similar
consideration holds for the spin-accumulation term. The origin of this
difference can be understood in the following way: we may visualize our wire
(or any system) as mapped onto an electric circuit which contains
energy-dependent conductors. In the case of purely elastic scattering we first
have to calculate the total conductance of the system for each energy. The
current is then found by averaging this spectral conductance with the
difference of distribution functions of the adjacent reservoirs. This
procedure yields Eq.~(\ref{eq:res-change}) for the change in the resistance.
In contrast, inelastic scattering equilibrates the local distribution of
electrons in a way that the chemical potential is equal to the potential found
from solving the circuit problem of the corresponding electric circuit. Thus,
Eq.~(\ref{RFSlinINEL}) follows the Kirchhoff's laws for our system. As we will
demonstrate below this difference can have significant consequences for the
temperature dependence of the resistance.

Let us now illustrate the temperature dependence of the linear response
conductance in the case of a metallic contact with an interface conductance
much larger than the conductance of the ferromagnetic wire. The total
conductance at sufficiently low temperatures is then $\langle
G_{\text{QP}}(\epsilon)+G_{\text{A}}(\epsilon)\rangle \approx G_{\text{A}}(0)
= 2G_{\text{BN}}$ and the quasiparticle conductance is $\langle
G_{\text{QP}}(\epsilon)\rangle \approx (8\pi k_{\text{B}}T/\Delta)^{1/2}\exp
(-\Delta/k_{\text{B}}T)$. The temperature must then be so low that
\begin{equation}
  \label{treshold}
  k_{\text{B}}T \lesssim \frac{\Delta}{
    \ln\left(R_{\text{BN}}/R_{\text{sf}}\right)}
\end{equation}
in order to prevent thermally assisted spin-current into the
superconductor.
\begin{figure}
  \includegraphics[width=7cm]{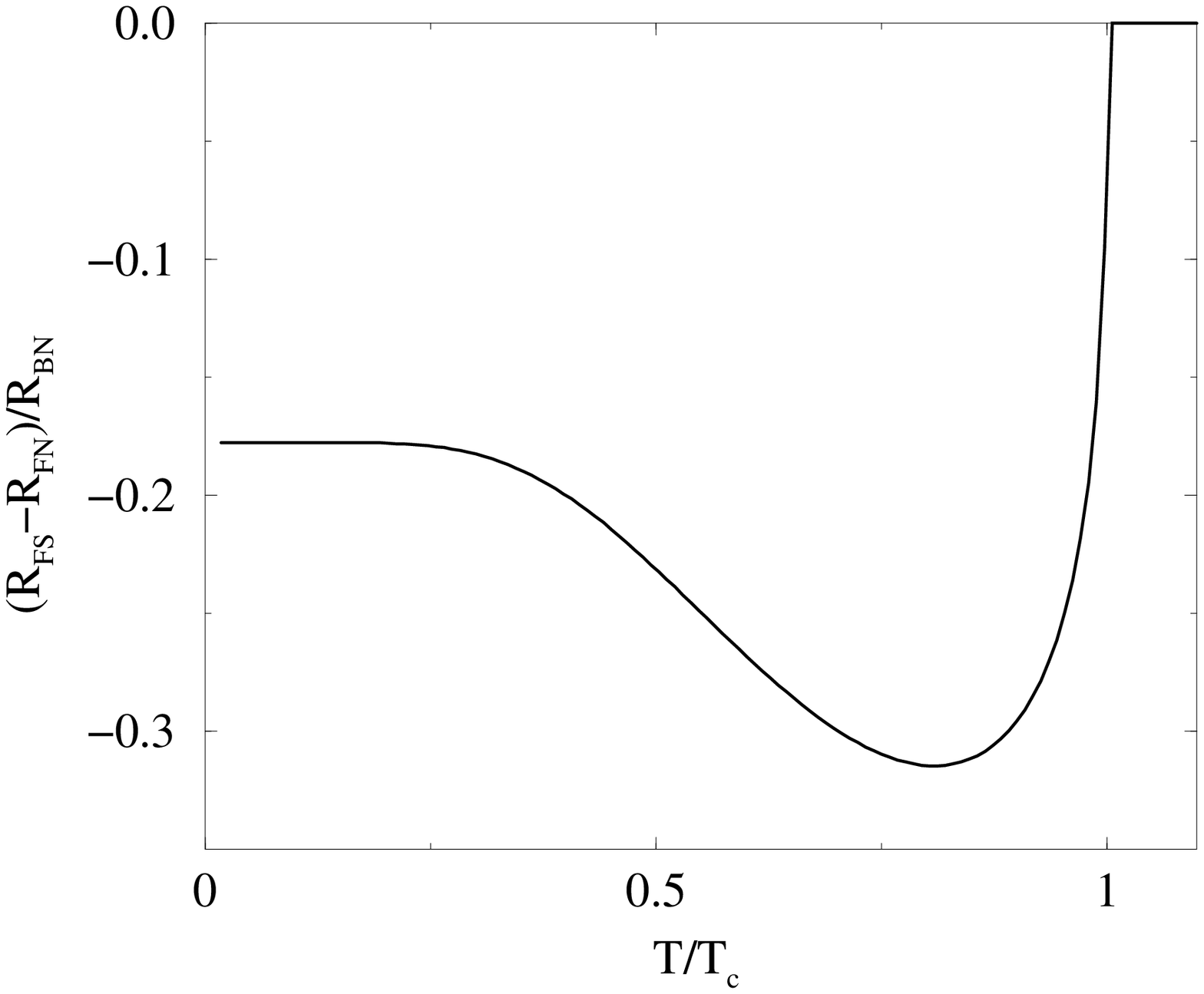}
  \caption{The ratio of the resistance $R_{\text{FS}}(T)$ to the
  interface resistance $R_{\text{BN}}$ as a function of the reservoir
  temperature $T$. The ferromagnetic wire and is parameterized by the
  polarization $\gamma=0.3$, the spin-flip diffusion length
  $l_{\text{sf}}=0.2 L$ and the interface resistance is
  $R_{\text{BN}}=0.05 R_{\text{F}}$. The spin-flip resistance is
  $R_{\text{sf}}=0.2 R_{\text{F}}$.}
\label{fig:LinIn}
\end{figure}

We show in Fig.\ \ref{fig:LinIn} the ratio of the linear response resistance
change $\Delta R_{\text{FS}}$ to the interface resistance $R_{\text{BN}}$ as a
function of the temperature $T$ for a metallic interface with
$R_{\text{BN}}=0.05 R_{\text{F}}$, polarization $\gamma =0.3$ and spin-flip
diffusion length $l_{\text{sf}}/L=0.2$. For these parameters we have a
`spin-flip' resistance corresponding to $R_{\text{sf}}=0.2 R_{\text{F}}$.  The
change in resistance below the superconducting transition temperature is due
to a competition between the excess resistance caused by the spin-flip
relaxation and the reduced interface resistance caused by Andreev reflection.
At $T=0$ we find from the approximate result (\ref{RFSlinINEL}) that
$R_{\text{FS}}-R_{\text{FN}}=0.5 R_{\text{BN}} - \gamma^2
R_{\text{sf}}=-0.14R_{\text{FN}}$ rougly corresponding to the numerical value
which has been obtained without making the approximation $\gamma^2 \ll 1$ and
$R_{\text{F}} \gg R_{\text{BN}}$.  Using the condition (\ref{treshold}) we
find that the spin-accumulation is strongly reduced around
$T/T_{\text{c}}=0.7$ and consequently the resistance of the system {\em
  decreases} before increasing again around $T/T_{\text{c}}=1$ where the
boundary resistance is increased. This explains the non-monotonic behavior of
the linear response resistance as a function of the temperature.

\subsection{Inelastic scattering - nonlinear response}

At a finite bias voltage the electron heating effects have to be taken into
account and the coupled equations for the electron temperature, the
spin-dependent chemical potentials (\ref{curcons}), (\ref{spinrelax}),
(\ref{encur}), and (\ref{encons}) and the boundary conditions
(\ref{eq:spin-currents}) and (\ref{eq:boundfer}) have to be solved
numerically.

\begin{figure}
  \includegraphics[width=7cm]{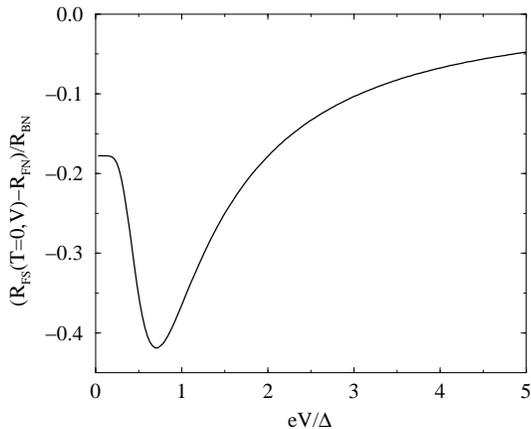}
  \caption{The ratio of the resistance $R_{\text{FS}}(T=0,V)$ to the
  interface resistance $R_{\text{BN}}$ as a function of the bias
  voltage $V$. The ferromagnetic wire and is parameterized by the
  polarization $\gamma=0.3$, the spin-flip diffusion length
  $l_{\text{sf}}=0.2 L$ and the interface resistance is
  $R_{\text{BN}}=0.05 R_{\text{F}}$. The spin-flip resistance is
  $R_{\text{sf}}=0.2R_{\text{F}}$.}
\label{fig:NonIn1}
\end{figure}

First let us discuss the transport properties when the electron-phonon
interaction is weak so that we have perfect conservation of energy current and
the left hand side of (\ref{encons}) can be set to zero. From the discussions
in the previous section we understand that there will be a reduction in the
excess resistance due to the spin-accumulation when the electron temperature
on the ferromagnetic side reaches condition (\ref{treshold}) so that there is
a significant spin-current entering the superconductor.  Roughly speaking, the
electron temperature on the ferromagnetic side is proportional to the applied
source drain bias.  Thus, as a crude approximation, we expect that the excess
resistance due to the spin-accumulation is lowered when
\begin{equation}
  eV \lesssim \frac{\Delta}{
  \ln\left(R_{\text{BN}}/R_{\text{sf}}\right)}\,.
\label{Vtreshold}
\end{equation}
We show in Fig.\ \ref{fig:NonIn1} the resistance change
$R_{\text{FS}}(T=0,V)-R_{\text{BN}}$ (\ref{diffres}) normalized by the
interface resistance $R_{\text{BN}}$ as a function of the bias voltage $V$.
As before, the interface resistance is $R_{\text{BN}}=0.05 R_{\text{F}}$, the
polarization $\gamma =0.3$ and the spin-flip diffusion length
$l_{\text{sf}}/L=0.2$. For these parameters we have a `spin-flip' resistance
corresponding to $R_{\text{sf}}=0.2 R_{\text{F}}$.  The change in resistance
below the superconducting gap is due to a competition between the excess
resistance caused by the spin-flip relaxation and the reduced interface caused
by the Andreev reflection. A dip in the resistance is seen around
$V=0.7\Delta$ which is correctly described by (\ref{Vtreshold}).  Below this
bias voltage the resistance is caused by the competetion between
spin-accumulation which enhances the resistance and the effective interface
resistance which reduces the resistance. At higher voltages the resistance is
only caused by the effective interface resistance and the reduction of the
resistance change $R_{\text{FS}}(T=0,V)-R_{\text{BN}}$ as a function of the
bias voltage is small.

In the limit of strong electron-phonon interaction the electron temperature
equals the lattice temperature. The spin-current into the superconductor is
then not enhanced due to thermal activation and consequently the
spin-accumulation on the ferromagnetic side is only reduced when the potential
on the ferromagnetic side of the interface is higher than the superconducting
gap. This occurs when
\begin{equation}
eV=\Delta \left(1+2\frac{R_{\text{F}}}{R_{\text{BN}}}\right)\,,
\label{cross}
\end{equation}
and thus at a potential that is much larger than the superconducting
gap, in contrast to the case of weak electron-phonon interaction.

In the intermediate regime the electron-phonon interaction should be included.
In order to illustrate the main physics we consider the case of a weak
polarization and set $\gamma=0$ and consequently there are no effects due to
spin-accumulation and the resistance change of the wire is only due to the
change of the effective boundary resistance. The chemical potential in the
ferromagnetic wire is thus spin-independent. Furthermore we consider the case
that the lattice temperature is zero, $T=0$ so that the electron temperature
arises solely due to electron heating.  Solving the diffusion equation
(\ref{curcons}) on the ferromagnetic side of the interface gives $\mu
(x)=-eVx/L + \mu(0) \left[1+x/L \right]$, where $eV$ is the applied bias and
$\mu(0)$ is the potential drop across the ferromagnet-superconductor
interface. The superconducting energy gap $\Delta$ presents a natural energy
scale for the problem.  We will characterize the strength of electron-photon
energy exchange by a dimensionless constant $\kappa=AL \zeta e^2 \Delta^3 /
G_{\text{F}}$, $\zeta$ is defined by the relation (\ref{enrel}).  The energy
diffusion equation then simplifies to
\begin{eqnarray}
  \frac{\pi^2}{6} \left(L_T \partial_x \right)^2
(k_{\text{B}}T_{\text{el}}/\Delta)^2
  & = &
  (k_{\text{B}}T_{\text{el}}/\Delta)^5 \nonumber \\ && -
\left((eV-\mu(0))/\Delta
\right)^2/\kappa \, ,
\label{endiff}
\end{eqnarray}
where we introduce a typical length-scale for the energy exchange
$L_T=L/\sqrt{\kappa}$.  If $\kappa \ll 1$ $L \ll L_T$ and the exchange is not
effective.  For longer wires, $\kappa$ becomes bigger than unity.  In this
case, the electron temperature develops a constant plateau in the
ferromagnetic wire and only changes rapidly within the length-scale $L_T$ near
the end-points $x=-L$ and $x=0$. It follows from (\ref{endiff}) that in this
case the temperature in the middle of the ferromagnetic wire becomes
\begin{equation}
\frac{k_{\text{B}} T_{\text{el}}}{\Delta} = \kappa^{-1/5}
\left(\frac{eV-\mu(0)}{\Delta}\right)^{2/5} \, .
\label{tempmid}
\end{equation}

\begin{figure}
  \begin{center}
    \includegraphics[width=7cm]{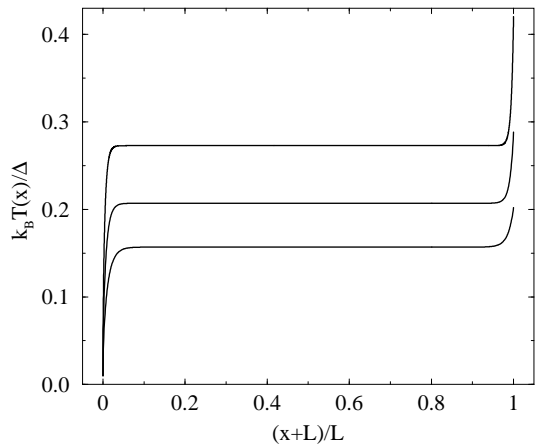}
    \caption{The spatially dependent electron temperature in the
      ferromagnetic wire. The upper curve is for a bias voltage
      $V=40\Delta$, the mid curve for a bias voltage $V=20\Delta$ and
      the lower curve for a bias voltage $V=10\Delta$. The metallic
      interface resistance is $R_{\text{BN}}=0.05R_{\text{F}}$, the
      reservoir temperature $T=0$ and the electron-phonon interaction
      strength $\kappa=10^6$.}
    \label{fig:temp}
  \end{center}
\end{figure}

We will now present numerical results of the temperature profile in the
ferromagnetic wire and the resulting resistance change using $\gamma=0$, a
metallic interface $R_{\text{BN}}/R_{\text{F}}=0.05$ for various values of the
electron-phonon coupling interaction. We show in Fig.\ \ref{fig:temp} the
spatially dependent electron temperature in the ferromagnetic wire for
$\kappa=10^6$ at a bias voltage $eV=40 \Delta$ (upper curve), $eV=20\Delta$
(mid curve) and $eV=10\Delta$ (lower curve). The electron temperature in the
middle of the wire follows from (\ref{tempmid}). There are rapid changes of
the electron temperature close to the ferromagnetic and superconducting
reservoirs and the temperature in the middle of the wire is lower than the
electron temperature close to the superconductor. The latter temperature is
important for the effective interface resistance.

\begin{figure}
  \begin{center}
    \centerline{\includegraphics[width=7cm]{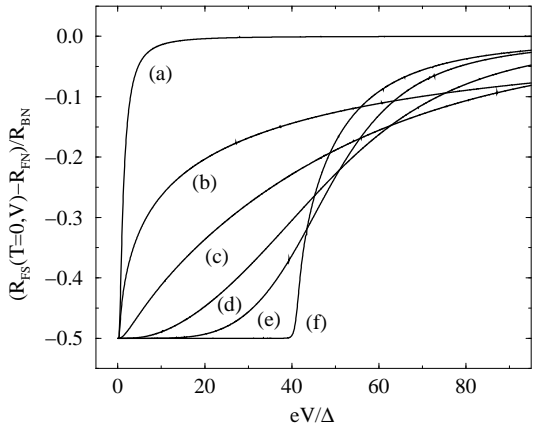}}
    \caption{Resistance change
      $(R_{\text{FS}}(T=0,V)-R_{\text{FN}})/R_{\text{FN}}$ as a function
      of the bias voltage $V$. The ferromagnetic wire is described by
      the polarization $\gamma = 0$, and the interface resistance is
      $R_{\text{BN}} = 0.05R_{\text{F}}$. Curve (a) corresponds to no
      electron-phonon interaction $\kappa = 0$. Curve (b), (c), (d), and
      (e) correspond to intermediate electron-phonon interaction,
      $\kappa = 10^2$, $\kappa = 10^4$, $\kappa = 10^6$ and $\kappa =
      10^8$, respectively. Curve (f) is for the case of strong
      electron-phonon interaction $\kappa = \infty$.  }
    \label{fig:NonIn2}
  \end{center}
\end{figure}

We show in Fig.\ \ref{fig:NonIn2} the resistance change as a function of the
bias voltage. The different solid lines show the current for different ratios
of the electron-phonon coupling starting from no electron-phonon interaction
(a) $\kappa=0$ going through intermediate electron-phonon interaction (b)
$\kappa=10^2$, (c) $\kappa=10^6$, (d) $\kappa=10^8$ to strong electron-phonon
interaction (f) $\kappa = \infty$ when the electron temperature equals the
lattice temperature, {\em e.g.} when there is no energy transfer between the
electron and phonon system. The cross-over bias voltage for the excess
resistance is sensitive to the strength of the electron-phonon interaction and
occurs from around $\Delta$ (a) to around $40 \Delta$ (f) (according to
(\ref{cross})). The dependence on the electron-phonon interaction parameter
$\kappa$ is rather weak as can be understood from (\ref{tempmid}). The local
electron temperature in the middle of the ferromagnetic wire is proportional
to $\kappa^{-1/5}$ and thus only has a very weak dependence on $\kappa$.

\section{Discussion of experiments}
\label{sec:discussion}

In this section we discuss the connection of our results with experiments of
Petrashov {\em et al.} \cite{Petrashov99:3281} and Giroud {\em et al.}
\cite{Giroud98:R11872}. It will turn out that most of the experimental results
can be understood on the basis of our calculations. Both experimental
arrangements we will discuss below contain F-S junctions where the
superconductor and the ferromagnet overlap in a certain region. The current
redistribution in these junctions will play an important role in the
following. Let us therefore introduce parameters characterizing these
juntions: the resistance of the interface is called $R_{\text{BN}}$ in
accordance with our previous consideration. Additionally $R_{\text{SJ}}$ will
be the resistance of the superconducting part of the overlap junction in the
normal state and $R_{\text{FJ}}$ the resistance of the ferromagnetic part of
the overlap junction.

In the experiment by Giroud {\em et al.} \cite{Giroud98:R11872} a
non-monotonic behaviour of the resistance below the superconducting critical
temperature was observed. The sample consisted of a ferromagnetic wire, the
resistance of which was measured in a 4-point arrangement. At some point a
superconducting strip was on top of the wire. In a second sample two such
strips were present and the resulting resistance change was twice as big as in
the case of one strip. Since our formulation is based on a single interface
and no coherent coupling between the two superconducting strips was found
experimentally, we concentrate here on the sample with one strip. The
resistance change in the two-strip sample is then simply twice that for the
single strip sample. The experimental arrangement is such that in the region
of the strip the current is redistributed among the ferromagnet and the
superconductor. In Appendix \ref{app1} we introduce a simple
quasi-one-dimensional model to calculate the effective resistance, these
results being used for comparison with experiment.  The resistance of the
superconducting Al-strip is $0.4\Omega$, the resistance of the ferromagnetic
part below the strip is $10\Omega$, and the resistance of the interface is
estimated to be $0.1\Omega$. Since the measured resistance change of the
F-wire shows no signature of the vanishing of the resistance of the
superconducting part, we believe that the real interface resistance is higher
than estimated in Ref.~\onlinecite{Giroud98:R11872}, in particular higher than
$R_{\text{SJ}}$.  This yields a total resistance of $R_{\text{eff}} =
2(R_{\text{FJ}}R_{\text{BN}})^{1/2}$, which is approximately of the order of a
few $\Omega$. A resistance change of the interface resistance $\Delta
R_{\text{B}}$ will than lead to an change of the effective resistance $\Delta
R_{\text{eff}}= (R_{\text{FJ}}/R_{\text{BN}})^{1/2}\Delta R_{\text{B}}$, which
in the case $R_{\text{FJ}}>R_{\text{BN}}$ is large than the resistance change
of the interface resistance itself. For the experimental values we have
$(R_{\text{FJ}}/R_{\text{BN}})^{1/2}\approx 10$ and thus a resistance change of
$\approx 0.2\Omega$, as observed in the experiment may result from a change of
the interface resistance $ R_{\text{BN}}\approx 0.1\Omega$ by $20\%$.

The results of Petrashov {\em et al.}~\cite{Petrashov99:3281} are more
intriguing, since the magnitude of the measured resistance drop in some of the
samples seems to be far too large to be explained without a `long-range'
proximity effect in the ferromagnet. We will concentrate here on three of the
four samples discussed in Ref.~\onlinecite{Petrashov99:3281}. In these samples
the transport through a long ferromagnetic wire with one ferromagnetic and one
superconducting contact is studied, this is in contrast to experiments of
Ref.~\onlinecite{Giroud98:R11872}.  The geometry is such that the
superconducting contact overlaps the ferromagnetic wire at one end and the
current has to pass through a tiny piece of the superconductor.  The three
samples differ in the interface resistance. Two samples with a low interface
resistance show large drops of the resistance of the order of $8\Omega$
respectively $16\Omega$ below the superconducting critical temperature. The
third sample has a higher interface resistance ($R_{\text{BN}}=41\Omega$) and
shows a small resistance increase of the order of $1.5\Omega$.

This agrees qualitatively with the results of our model. Indeed, the bigger
resistance of the boundary usually means a formation of a ticker tunnel
barrier such that the transmission eigenvalues are shifted towards zero.  Our
model does predict a resistance decrease for a fairly transparent interface
and changes to an increase for a more tunnel-like interface. This is shown
{\em e.g.} in Fig.~\ref{fig:el-spinaccu}.  However, quantitatively one would
expect that the resistance changes below the superconducting transition
temperature are {\em always} of the order of the boundary resistance itself.

This is obviously not the case in the experiment with
$R_{\text{BN}}=41\Omega$, where the measured resistance change is about 40
times smaller. The first idea is that the resistance drop of the samples with
better interface may possibly be accounted for by combining the effect of
current redistribution and the apparent enhancement of Andreev reflection
discussed in Sec.\ref{sec:results}.  Again we calculate the effective
interface resistance in a quasi-one-dimensional model (see Appendix
\ref{app1}).  In the limit of a small interface resistance $R_{\text{eff2}} =
(R_{\text{FJ}}R_{\text{BN}})^{1/2}$. A change of the interface resistance
again results in an apparently larger change of the effective resistance
$\Delta R_{\text{eff2}}= (R_{\text{FJ}}/R_{\text{BN}})^{1/2} \delta
R_{\text{B}}/2$. We may speculate that the large resistance drop observed in
the experiment by Petrashov {\em et al.}  can possibly be explained by this
effect together with the observation made in \ref{sec:results} that a spin
dependent interface may cause another apparent enhancement of the
Andreev-reflection.

\begin{figure}
  \begin{center}
    \includegraphics[width=7cm]{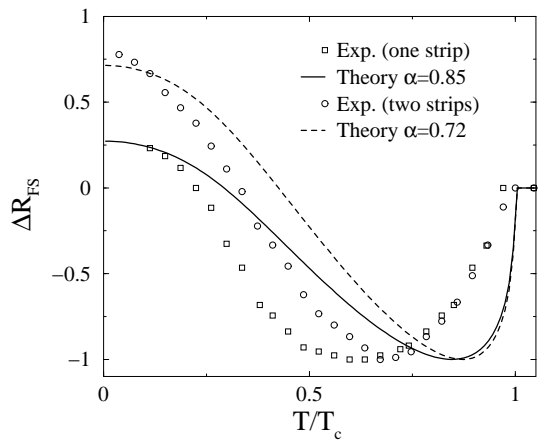}
    \caption{Comparison of experiment of Giroud {\em et al.}
      \cite{Giroud98:R11872} and theoretical calculations. The two
      experimental data sets are for the sample with one superconducting
      strip (squares) and for the sample with two superconducting strips
      (circles).  The theoretical curves are obtained from
      Eq.~(\ref{eq:gfs}) with the following parameters:
      $R_{\text{F}}=100 R_{\text{BN}}$, spin relaxation length
      $l_{\text{sf}}=0.012$, conductivity polarization $\gamma=0.3$ and
      a double interface contact. All curves are normalized to the
      respective maximal value.}
    \label{fig:exp-fit}
  \end{center}
\end{figure}

There may be a more radical explanation for a small relative resistance
change.  In fact, the morphology of the metal-ferromagnet interfaces has not
been yet sufficiently studied.  The actual structure of the interface may be
complicated.\cite{petrashov-priv} To illustrate how this can affect the
results let us consider a simplistic model of a double interface.  We
speculate that a thin layer of {\it magnetic} alloy separates the ferromagnet
and the superconductor. The boundary scattering then occurs in two stages: at
the ``inner" interface between the ferromagnet and the alloy and at the
``outer" interface between the superconductor and the alloy.  Since the
proximity effect is quenched in magnets, the resistance of the ``inner"
interface is not affected by the superconducting transition whereas the
resistance of the ``outer" interface acquires a change described above. This
leads to a smaller relative resistance change.

In Fig.~\ref{fig:exp-fit} we show a comparison between the experimental
results of Giroud {\em et al.} and our calculation for a contact with a
transmission eigenvalue distribution of the model just described (see Appendix
\ref{sec:app2}). The maximal relative resistance change of the contact is
$\approx 16\%$ in our calculation. According to above considerations this may
be enhanced to a measured resistance change of the order of $\approx
(R_{\text{FJ}}/R_{\text{BN}})^{1/2} 16\% \delta R_{\text{B}}\approx 160\%
\delta R_{\text{B}}$ in agreement with measurements.

\section{Conclusions}
\label{sec:concl}

To summarize, we have calculated the resistance of a diffusive ferromagnetic
wire in contact with a superconducting reservoir in the linear and non-linear
regime with purely elastic and inelastic scattering. It has been demonstrated
that most of the recent experimental results can be understood in the absence
of a superconducting proximity effect in the ferromagnet. Spin accumulation
leads to an enhanced resistance below the superconducting transition
temperature whereas Andreev reflection can lead to a decreased resistance
below the superconducting transition temperature. The competition between
these two mechanism determines the sign of the resistance change. The
magnitude of the resistance change is of the order of the interface resistance
or the spin-relaxation resistance.  Electron heating can dramatically modify
the nonlinear response resistance and change the bias dependence by orders of
a magnitude.

\acknowledgements This work was partially supported by the ``Stichting voor
Fundamenteel Onderzoek der Materie'' (FOM) and a Feodor Lynen Fellowship of
the ``Alexander von Humboldt-Stiftung'' (W.~B.) and the Norwegian Research
Council (A.~B.).

\appendix

\section{Current redistribution in an overlap junction}
\label{app1}

\begin{figure}[htbp]
  \begin{center}
    \includegraphics[width=5cm]{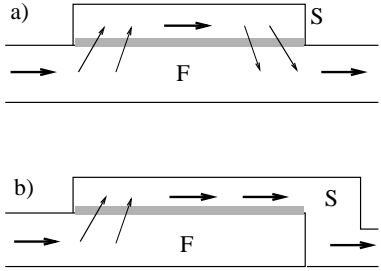}
  \caption{
    Schematic picture of current redistribution in an overlap
    junction.  The geometry a) corresponds to the experiment by Giroud
    {\em et al.}  \cite{Giroud98:R11872}. If the resistance of the
    ferromagnet below the overlap region is the highest, a
    considerable part of the current flows through the interface and
    the superconductor. The geometry in b) corresponds to the geometry
    of Petrashov {\em et al.}  \cite{Petrashov99:3281}. Here the
    current is forced to leave the contact region through the
    superconductor.  }
    \label{fig:current}
  \end{center}
\end{figure}

Here we introduce a quasi-one-dimensional model to account for the
redistribution of the current under an overlap junction as used in the
experiments. The two geometries we have in mind are depicted in
Fig.~\ref{fig:current}.

To estimate the measured resistance we use the following
quasi-one-dimensional model for the current redistribution in the overlap
region of length $d$.  The currents $I_F(x)$ in F and $I_S(x)$ in S in
direction of the ferromagnetic wire follow from Ohm's law
\begin{eqnarray}
  \label{eq:ohmlaws}
  I_{\text F}(x) =
  \frac{d}{R_{\text{FJ}}}\frac{d U_{\text{F}}(x)}{dx}
  &\quad,\quad&
  I_{\text S}(x)  =
  \frac{d}{R_{\text{SJ}}}\frac{d U_{\text{S}}(x)}{dx}\,,
\end{eqnarray}
where $R_{\text{FJ(SJ)}}$ is the resistance of the ferromagnetic part under
(superconducting part above) the contact and $U_{F(S)}$ the respective
voltage.  Current conservation dictates
\begin{equation}
  \label{eq:current-conserv}
  \frac{d I_{\text F}(x)}{dx}=
  -\frac{d I_{\text S}(x)}{dx}=
  \frac{U_S(x)-U_F(x)}{R_BN d}\,,
\end{equation}
is the current per unit length through the contact resistance is
$R_{BN}$. Boundary conditions are obviously that the total voltage drop is
equal to $V$ and no current leaves the system through the boundary to
vacuum.

Solving these equations for the geometry of Giroud {\em et al.}
\cite{Giroud98:R11872} we find the effective resistance of this part to be
\begin{eqnarray}
  \label{eq:Reff-giroud}
  R_{\text{eff}}&=&\frac{R_{\text{FJ}}}{R_{\text{FJ}}+R_{\text SJ}}
  \\\nonumber &&\times
  \left(R_{\text{SJ}}+
    \sqrt{\frac{4R_{\text{BN}}}{R_{\text{SJ}}+R_{\text{FJ}}}}
    \tanh\left(
      \sqrt{\frac{R_{\text{FJ}}+R_{\text{SJ}}}{4R_{\text{BN}}}}\right)\right)
    \,.
\end{eqnarray}
Of specific interest if the case that $R_{\text{FJ}}\gg R_{\text{BN}} \gg
R_{\text{SJ}}$ in which case we obtain $R_{\text{eff}}\approx
2(R_{\text{FJ}}R_{\text{BN}})^{1/2}$.

A calculation similar to the previous for the geometry of Petrashov {\em et
  al.}\cite{Petrashov99:3281} leads to an effective measured resistance of
\begin{equation}
  \label{eq:reff2}
  R_{\text{eff2}}=\frac{\sqrt{R_{\text{FJ}}R_{\text{BN}}}}{
    \tanh\sqrt{R_{\text{FJ}}/R_{\text{BN}}}}
\end{equation}
in the limit of vanishing resistance of the superconductor on top of the
ferromagnet. The difference to the previous calculation is that here the
current enters the junction through the ferromagnet, but has to leave the
junction through the superconductor.

\section{Transmission eigenvalues of a double interface}
\label{sec:app2}

\begin{figure}[htbp]
  \begin{center}
    \includegraphics[width=7cm]{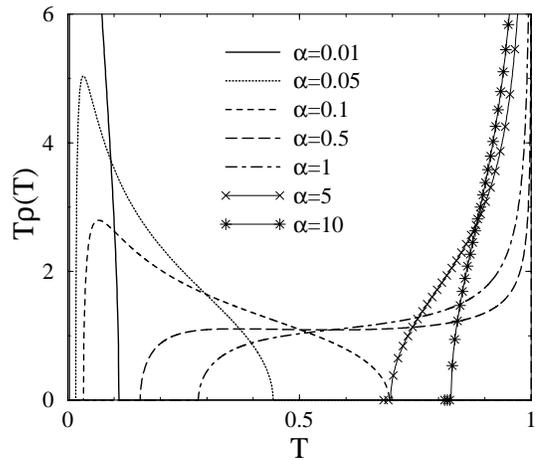}
  \caption{
    Transmission eigenvalue distribution for the double interface. The
    ratio $\alpha=G_{\text{T}}/2G_{\text{QPC}}$ is varied between the
    different curves.}
  \label{fig:trans-distr}
\end{center}
\end{figure}

The distribution of transmission eigenvalues of a double interface as
described in the text can be found with the technique described in
Ref.~\onlinecite{nazarov:94}. For details we refer to these articles. We model
the double interface by a ballistic contact and a tunneling barrier in series.
The tunneling barrier of conductance $G_{\text{T}}$ models the sharp drop in
the potential due to the band structure mismatch, whereas the region close to
that interface is treated as a collection of unit transmission channels with a
total conductance $G_{\text{QPC}}$ The distribution of transmissions can be
found from the solution of
\begin{equation}
  \label{eq:i-phi}
  I(\Phi)=G_{\text{T}} \sin(\Phi-\theta)
  =2G_{\text{QPC}}\tan\left(\frac{\theta}{2}\right)\,.
\end{equation}
The distribution of transmission eigenvalues $\rho(T)$ is found by analytic
continuation into the complex plane
\begin{equation}
  \label{eq:distr}
  \rho(T)=\frac{1}{e^2}\frac{1}{T\sqrt{1-T}} \text{Re}
  \left[ I\left(\pi+2 i \,\text{acosh}\frac{1}{\sqrt{T}}\right) \right]\;.
\end{equation}
The dependence on the two separate conductances may be eliminated in favor of
the total conductance of the contact
$G_{\text{BN}}=G_{\text{T}}G_{\text{QPC}}/(G_{\text{T}}+G_{\text{QPC}})$ and
the ratio of the two $\alpha=G_{\text{T}}/2G_{\text{QPC}}$. The transmission
eigenvalue distribution then only depends on $\alpha$. It is plotted in
Fig.~\ref{fig:trans-distr} for several values of $\alpha$. For small values of
$\alpha$ the contact is dominated by the tunnel barrier resulting in a shift
of the transmission eigenvalues to lower values and a gap above a certain $T$.
Higher $\alpha$ shift the distribution to larger transmission eigenvalues and
a gap opens up for low transmission eigenvalues. For a range $0.1\lesssim
\alpha\lesssim 0.5$ the distribution restricted to a finite interval of
transmission eigevalues. At even higher values of $\alpha$ the upper gap
closes and the distribution becomes more and more peaked at $T=1$.

\end{multicols}

\begin{references}
\bibitem{delta} In the actual calculation we have approximated the temperature
  dependence of the superconducting gap by $\Delta(T)=1.76 T_{\text{c}}
  \tanh\left(1.74(T_{\text{c}}/T-1)^{1/2}\right)$.
\end{references}
\end{document}